\documentstyle[graphicx]{mn}

\title{The Contribution of Faint Blue Galaxies to the Sub-mm Counts and
Background}
\author[Busswell et al.]
{
Geoff S. Busswell, Tom Shanks \\ 
Department of Physics, University of Durham, Science Labs, South Road, Durham
DH1 3LE, UK}

\begin{document}

\maketitle

\begin{abstract}
Observations in the submillimetre waveband have recently revealed a new
population of luminous, sub-mm sources.  These are proposed to lie at
high redshift and to be optically faint due to their high
intrinsic dust obscuration.   The presence of dust has been previously 
invoked in 
optical galaxy count models which assume $\tau=9$ Gyr Bruzual \& Charlot 
evolution for spirals and these fit the count data well from U to K.
We now show that by using either a 1/$\lambda$
or Calzetti absorption law
for the dust and re-distributing the evolved spiral galaxy UV
 radiation into the
far infra-red(FIR), these models can account for all of the `faint'($\leq1$mJy)
$850\mu$m galaxy counts, but fail to fit 'bright'($\ge2$mJy) sources,
indicating that another explanation for the sub-mm counts may apply at
brighter fluxes(e.g. QSOs, ULIRGs). We find
that the main contribution to the faint, sub-mm number counts is in the
redshift range $0.5 < z < 3$, peaking at $z\approx 1.8$.  The above model, 
using either dust law,
 can also explain a
significant proportion of the extra-galactic background at $850\mu$m as
well as producing a reasonable fit to the bright $60\mu m$ IRAS counts.

\end{abstract}


\section{Introduction}
 
The SCUBA camera \cite{holland99} on the James Clerk Maxwell Telescope
has transformed our knowledge of dusty galaxies in the distant Universe
as a result of the discovery of a new population of luminous, dusty,
infra-red galaxies (Smail et al. 1997; Ivison et al 1998).  It has been 
proposed that 
these galaxies may be similar to IRAS
ULIRGs (ultra-luminous infra-red galaxies) 
which appear to be starbursting/AGN galaxies, containing large amounts
of dust.  The possibility that much star-formation is hidden by dust 
means that sub-mm observations can give an invaluable insight
into the star-formation history of the Universe.  This view aided by
the redshifting
of the thermal dust  emission peak in starbursting galaxies into the FIR, which
results in a negative k-correction in the sub-mm. By this route, we can
therefore study our Universe all the way back to very early times and
gain unprecedented insight into the formation and evolution of galaxies.

The first sub-mm galaxy to be detected by SCUBA was SMM J02399-0136
\cite{ivison98}, which is a massive starburst/AGN at
z=2.8 and the current situation is
that the complete 850$\mu$m sample from all the various groups
consists of well over 50  sources (Blain et al. 1999; Eales et al. 1999; 
Hughes et al. 1998; Holland et al. 1998; Barger et al. 1998; Smail
et al. 1997).  Optical and near
infra-red(NIR) counterparts have been identified for about a third of the
sources, although the reliability of these
identifications varies greatly.  This problem is due to the fact that 
the $\approx 15''$ FWHM of the SCUBA beam results in $_-^+3''$ 
positional errors
on a sub-mm source, so there is a reasonable chance that several 
candidates could lie within these errors.  Also, there is no guarantee
that the true source will be detected down to the optical flux limit as, for 
example, many of the sources have been shown to be very red objects 
(Dey et al. 1999; Smail et al. 1999: Ivison et al. 2000) and
therefore have not been found in optical searches for sub-mm sources. 

What has proved extremely
 enlightening is that radio counterparts at 1.4GHz have 
now been identified
for many of the sub-mm sources (Smail et al. 2000: Ivison et al. 2000) 
providing much more accurate angular positions 
($<1''$ in some cases) and reasonably accurate photometric redshifts.  
Various 
groups have obtained redshift distributions of sub-mm samples
(Hughes et al. 1998; Barger et al. 1999a; Lilly et al. 1999; Smail et al. 2000) 
and they all 
derive results that are consistent with a mean redshift in the range 
$1 < z < 3$.  The fact that almost all of the sources are associated with 
mergers or interactions seems to confirm that the population of sources 
contributing at the `bright' ($>2$mJy) sub-mm fluxes (since most of
the sources so far discovered are `bright') are similar to local
IRAS ULIRG's, ie massive, starbursting/AGN galaxies 
which are extremely luminous 
in the far-infra-red.
This hypothesis is strengthened further by the fact that the only two sub-mm 
sources (SMM J02399-0136 and SMM J14011+0252) with reliable redshifts have 
been 
followed up with millimeter wave 
observations (Frayer et al. 1998, 1999), resulting in CO 
emission being detected at the
redshifts of both sources (z=2.8 and z=2.6), a characteristic indicator 
of large quantities of molecular gas present in IRAS galaxies.

The nature of the fainter ($\le1$mJy) sub-mm population is, however, the focus
 of this paper.  
It has been claimed by Peacock et al. (2000) and Adelberger et al. (2000)
that the Lyman Break Galaxy(LBG) population could not only contribute
significantly to the faint sub-mm number counts, but could also account
for a substantial proportion of the background at $850\mu$m. 
This may indicate that ULIRG's 
cannot explain all of the sub-mm population and that the UV-selected galaxy
 population, which are predicted to be evolved spirals by the Bruzual \&
Charlot models, may in fact make a substantial contribution.
It is exactly this hypothesis our paper addresses.   

In this paper we will first review the situation regarding the optical
galaxy counts, focusing in particular on the models of Metcalfe et al. (1996).
  These simple models which use a $\tau =9$Gyr SFR for spirals
and include the effects of dust give good fits to galaxy counts and
colours from U to K.  The idea is then to see whether this
combination of exponential SFR and relatively small amounts of dust in the
 first instance
(A$_B=0.3$ mag. for the $1/\lambda$ law), which would re-radiate the spiral 
ultra-violet (UV)
radiation into the FIR, could cause a significant contribution to the
sub-mm galaxy number counts and background at $850\mu $m.  Our modelling
will be described in section 3 and then in section 4 our predicted
contribution to the $850\mu $m and $60\mu m$ galaxy counts and the
extra-galactic background in the sub-mm will be shown.  Also in this
section we demonstrate how to get a fit to the background in the
$100-300\mu m$ range by using warmer, optically-thicker dust in line
with that typically seen in ULIRG's. We will then discuss the
implications of our predictions in section 5 and conclude in section 6.

\section{The Optical Counts}

It is well known that non-evolving galaxy count models, where number
density and luminosity of galaxies remain constant with look-back
time, do not fit the optical number counts e.g. \cite{shanks84}, as there
is always a large excess of galaxies faintwards of $B\sim 22^m$.  One
way to account for this excess of 'faint blue galaxies' is to
investigate the way galaxy evolution will influence the optical number
counts.
Metcalfe et al (1996) showed that by assuming that the number density of
galaxies remains constant, the Bruzual and Charlot(1993) evolutionary
models of spiral galaxies with a $\tau = 9$Gyr SFR give excellent fits
to the optical counts.  The galaxy number counts are normalised at
$B\sim 18^m$ so that the non-evolving models give good fits to the $B$
band data and redshift distributions in the range $18^m < B < 22.^m5$.
With this high normalisation, the models of the galaxy counts
represent both spiral and early-type galaxies extremely well for
$17^m < I < 22^m$ (Glazebrook et al. 1995a, Driver et al 1995) and 
also the less steep $H/K$ counts
out to $K\sim 20^m$.  The evolution model then produces a reasonable
fit to the fainter counts to $B\sim 27^m$, $I\sim 26^m$, $H\sim28^m$.


Metcalfe et al (1996) included a $1/\lambda$ internal dust absorption law with
$A_B=0.3$ for spirals to prevent the $\tau=9$ Gyr SFR from over-predicting the
numbers of high redshift galaxies detected in faint B$<24$ redshift surveys 
(Cowie et al 1995).  This $1/\lambda$ dust law differs from the Calzetti(1997)
dust law derived for  starburst galaxies, in that for a given $A_B$, more
radiation is absorbed in the UV. The Calzetti dust law is used by Steidel et
al(1999) to model their `Lyman Break' galaxies; they find an average
E(B-V)=0.15 which gives $A_B=0.87$mag and $A_{1500}=1.7$mag. 
This compares to our
$A_{1500}=0.9$mag with $A_B=0.3$mag. 
Both models also
fail to predict as red colours as observed for  the U-B colours of spirals in
the Herschel Deep Field (Metcalfe et al 1996).  However, if we assumed
E(B-V)=0.15 for our z=0 spirals, as compared to our E(B-V)=0.05, then the rest
colours of spirals as predicted by the Bruzual \& Charlot model might start to
look too red as compared to what is observed. Otherwise, the main difference
between these two dust laws is that the Calzetti law would produce more overall
absorption and hence a higher FIR flux from the faint blue galaxies. Thus in
some ways our first use of the $1/\lambda$ law appears
 conservative in terms of the
predicting the faint blue galaxy FIR flux.  Later, we shall experiment by
replacing the $1/\lambda$ law with the Calzetti(1997) law in our model.

So this pure luminosity evolution (PLE) model with $1/\lambda$
dust and $q_0=0.05$ then slightly under-estimates the faintest optical  counts
but otherwise fits the data well, whereas for $q_0=0.5$ the underestimate (with
or without dust) is far more striking.  An extra population of galaxies has to
be invoked at high redshift to attempt to explain this more serious discrepancy
for the high $q_0$ model.  This new population was postulated to have a
constant SFR from their formation redshift until $z\sim 1$ and then the Bruzual
\& Charlot models predict a dimming of $\sim 5^m$ in $B$ to form a red dwarf
elliptical (dE) by the present day and therefore has the form of a
'disappearing dwarf' model \cite{babrees92}. No dust was previously assumed in
the dE population but this assumption is somewhat arbitrary.

The $\tau=9$Gyr SFR was inconsistent with the early observations at low
redshift from Gallego et al.(1996)  and this is partly accounted for by the
high normalisation of the optical number counts at $B\sim 18^m$.  There is
still a problem with the UV estimates from the CFRS UV data of Lilly et al at
z=0.2. More recent estimates of the global SFR at low redshift based on the 
[OII]
line (Gronwall et al.1998; Tresse \& Maddox 1998; Hammer and Flores 1998)
indicates that the decline from z=1 to the present day may not be as sharp as
first thought and that the $\tau=9$Gyr SFR in fact provides a better fit to
this low redshift data. Metcalfe et al (2000) have further found 
that this model 
also agrees well with recent  estimates of the luminosity function of the z=3
Lyman break galaxies detected by Steidel et al.(1999).

The main question then that we will address in this paper is whether the
small amount of internal spiral  dust absorption assumed in these 
models which give an excellent fit to the optical galaxy counts,  could cause 
a significant contribution to the sub-mm number counts and background at $850
\mu$m.

\section{Modelling}


Using the optical B band parameters for spiral galaxies, we attempt to
predict the contribution to the sub-mm galaxy counts and background at
$850\mu m$ by using a 1/$\lambda$ absorption law for the dust and
re-radiating the spiral UV radiation into the FIR.  We use the Bruzual
\& Charlot(1993) galaxy evolution models with
$H_0=50$km$s^{-1}$Mpc$^{-1}$ and a $\tau$=9 Gyr SFR - with a galaxy
age of 16 Gyr in the low $q_0$ case, and 12.7 Gyr in the high $q_0$
case to produce our 1M$\odot$ galactic spectral energy
distribution(SED) as a function of redshift.  We then use the equation

\begin{equation}
G_{abs}(z) = \int F_{\lambda}(z)(1-10^{-0.4*A_B*(4500/\lambda)})d\lambda
\end{equation}
~

as used by Metcalfe et al(1996), which is used to calculate the
radiation absorbed by the dust, $G_{abs}$($ergss^{-1}$), for our
1M$\odot$ model spiral galaxy as a function of z, using our
$1/\lambda$ absorption law with A$_B=0.3$.  Since Bruzual \& Charlot
provides us with a 1M$\odot$ SED at each redshift increment, we need
to calculate the factor required to scale this SED (after the effect
of absorption from the dust) to obtain that of a galaxy with absolute
magnitude $M_B$ at zero redshift, and this factor will then remain
constant for $M_B$ galaxies at all other redshifts. This then provides
a zero point from which to calculate scaling factors for all the other
galaxies in our luminosity functions.  We find the scaling factor for
an $M_B$ galaxy by making use of a relation from Allen(1995)

\begin{equation}
m_B = -2.5log(\int B_{\lambda}{\it f}_{\lambda}d \lambda) - 12.97
\end{equation}
~
where ${\it f}_{\lambda}$ is the received flux($ergs^{-1}\AA^{-1}cm^{-2}$) and 
$B_{\lambda}$ is the B band filter function.  By re-arranging, setting
$m_B$=$M_B$ and then multiplying by $4\pi(10pc)^2$ we obtain the total
emitted power, $L_B$($ergs^{-1}$) in the B band from an $M_B$ galaxy

\begin{equation}
L_B=4\pi(10pc)^2.10^{[-0.4(M_B+12.97)]}
\end{equation}
~

The intensity emitted in the B band, after absorption by the dust from
our 1M$\odot$ galaxy, $L_{BM_{\odot}}$ is then calculated by
integrating the SED, assuming a flat B band filter, between $4000\AA$
and $5000\AA$.  

\begin{equation}
L_{BM_{\odot}}=\int
F_{\lambda}(z)10^{-0.4*A_B*(4500/\lambda)}B_{\lambda}d\lambda
\end{equation}
~

The scaling factor  to scale a Bruzual \& Charlot $1M_{\odot}$spectral 
energy distribution for a galaxy of absolute magnitude, M$_B$, is then 
defined by the ratio $L_B/L_{BM_{\odot}}$. 
    
The way the dust will re-radiate this absorbed flux depends on its
temperature, particle size and chemical composition.  However the
normalisation of the re-radiated flux from a galaxy with absolute
magnitude M$_B$, at redshift z, is already determined (the quantity
$G_{abs}$$E_B/E_{BM_{\odot}}$).  We will adopt a simple model by
assuming a mean interstellar dust temperature of 15K, \cite{bianchi99}
and also a modest warmer component of 45K, (the actual luminosity
ratio we use is $L_{45K}/L_{15K}=0.162$), which would come from
circumstellar dust \cite{dom99} and is needed in order to fit counts
at shorter wavelengths eg. $60\mu m$.  The effect of varying
the dust parameters is explored in section 4.  We then simply scale the Planck
function so that

\begin{equation} 
C(z,M_B) \int_{-\infty}^{\infty}\beta(\lambda, T)d\lambda  = 
G_{abs}L_B/L_{BM_{\odot}}
\end{equation}
~

where C(z,M$_B$) is the scaling factor, which is a function of z and
M$_B$, $\beta(\lambda, T)$ is the Planck function (in this case a sum
of two Planck functions) and $\kappa_d (\lambda) \propto
\lambda^{-\beta}$, where $\kappa_d (\lambda)$ is an opacity law (we
use $\beta=2.0$ for each Planck function to model optically thin
dust).

We then calculate the received $850\mu m$ flux, S(z,M$_B$), from a galaxy  
with absolute magnitude $M_B$ and redshift z using the equation

\begin{equation}
\ S(z,M_B)=\frac{C(z,M_B)  \lambda_e^{-\beta} \beta(\lambda_e, T)}
{4\pi (1+z) d_L^2} 
\end{equation}
~

where C(z,M$_B$) is defined from (4) and $\lambda_e$ is equal to $850
\mu m$/(1+z).  We can then obtain the number count of galaxies with
absolute magnitude between M$_B$ and M$_B+dM_B$ and redshift between z
and z+dz for which we measure the same flux density S(z,M$_B$) at
$850\mu m$ (see (4)).

\begin{equation}
dN(z,M_B) = \phi(M_B) \frac{dV}{dz}dM_Bdz
\end{equation}
~

where $\phi (M_B)$ is the optical Schechter function and
$\frac{dV}{dz}$ is the cosmological volume element. Then the integral
source counts N$(>S_{lim})$ are obtained, for each value of $S_{lim}$,
by integrating (5) over the range of values of $M_B$ and $z$ such that
S(z,M$_B)>S_{lim}$, where S(z,M$_B$) is defined in (4).
  

%

\begin{equation}
N(>S_{lim})=\int_{M_B}\int_{z}\phi(M_B)\frac{dV}{dz}dM_Bdz
\end{equation}
~

It is straightforward to then obtain model predictions of the FIR
background for a given wavelength.  The intensity, dI, at $850\mu m$
from galaxies with absolute magnitudes between $M_B$ and $M_B+dM_B$
and redshifts between z and z+dz is given by multiplying the number of
galaxies with these z's and M$_B$'s by the flux density which we would
measure from each

\begin{equation}
dI_{850}=S(z,M_B)\phi(M_B)\frac{dV}{dz}dM_Bdz
\end{equation}
~
and then we simply integrate over all absolute magnitudes and all redshifts
($0 < z < 4$ in this case)

\begin{equation}
I_{850} =  \int_{M_B}\int_{z} S(z,M_B)\phi(M_B)\frac{dV}{dz}dM_Bdz
\end{equation}
~

\section{Predictions}
\begin{figure}
\centering
\includegraphics[width=3in,totalheight=3in,angle=-90]{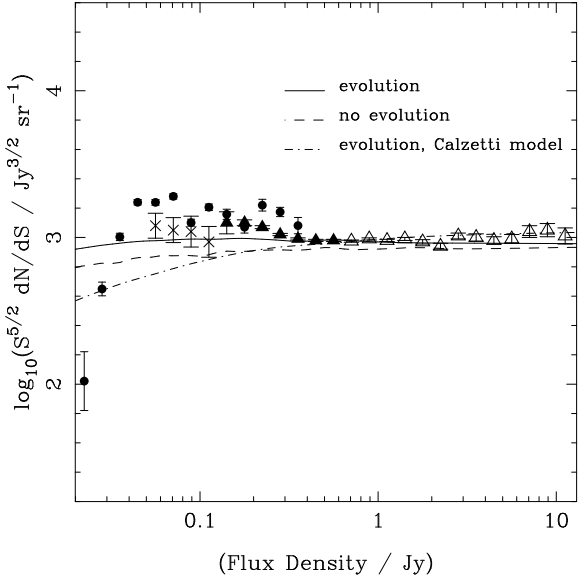}

\caption{The $60\mu m$ differential number counts.  The graph shows
the evolution and no-evolution models for a low $q_0$ Universe(the
corresponding high $q_0$ models are indistinguishable) along with the
observed $60\mu m$ counts of IRAS galaxies down to a flux limit of
0.6Jy, plotted in the format used by Oliver et al.(1992).  The crosses are from
Hacking \& Houck(1987), the empty triangles from Rowan-Robinson et al.
(1990), Saunders et al.(1990) are the filled triangles and the circles
are Gregorich et al.(1995) and Bertin et al.(1997).  We use a
two-component dust temperature of 15K and 45K to model both
interstellar and circumstellar dust respectively.  Other parameters
used are $\beta=2.0$, $H_0=50$ and a redshift of formation of $z=4$.
The dot-dashed line shows the same evolution model using the Calzetti dust
law with three dust temperature components of 15, 25, and 32K.  
This fits the IRAS counts less well at $<0.2$mJy, and this is because of
the lack of a 45K dust component meaning that there is much less
 thermal emission from
the dust at $60\mu$m}.
\label{fig:counts60}
\end{figure}

\begin{figure}
\centering
\includegraphics[width=3in,totalheight=3in,angle=-90]{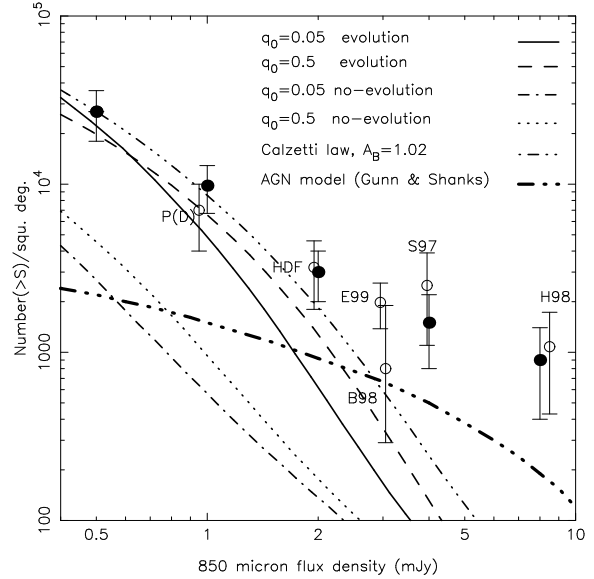}
\caption{The $850\mu$m integral number counts.  The filled circles
show the results of the SCUBA Lens Survey (Blain et al. 1999); the
open circles are as labelled:S97 - Smail, Ivison \& Blain(1997); B98 -
Barger et al.(1998); H98 - Holland et al.(1998); E99 - Eales et
al.(1999); HDF, P(D) - Hughes et al.(1998).  Also shown are our
predictions for $q=0.05$ and $q=0.5$ models with and without Bruzual
\& Charlot evolution, using the parameters from Fig. \ref{fig:counts60}.
Both the high and low $q_0$ models, with evolution (dashed and solid
curves), do very well with the faint counts but fail the most luminous
sources.  In the no evolution cases(dotted and dot-dashed) the high
$q_0$ model again predicts more galaxies then the low $q_0$ model, but
they both underpredict the faint $850\mu$m counts by about an order
of magnitude and then again fall away again at the higher flux
densities. The graph also shows a predicted contribution from AGN (Gunn \& 
Shanks 1999) and a model using the calzetti dust law (the two dot-dot-dot-
 dashed
 curves).  The AGN model (the steeper of the curves) predicts that, at most, 
QSO's could contribute 30 percent
of the background at $850\mu$m, and these models
do much better in the number counts at brighter fluxes, but they fail to 
contribute at the $0.5$mJy level where we predict that faint blue 
galaxies are  dominant.  Our Calzetti dust law uses three dust temperature
components (see Fig. \ref{fig:counts60}, and as
with our $1/\lambda$ dust law, it can account for the faint number counts
but then fails the much brighter sources. }
\label{fig:counts}
\end{figure}

\begin{figure}
\centering
\includegraphics[width=3in,totalheight=3in,angle=-90]{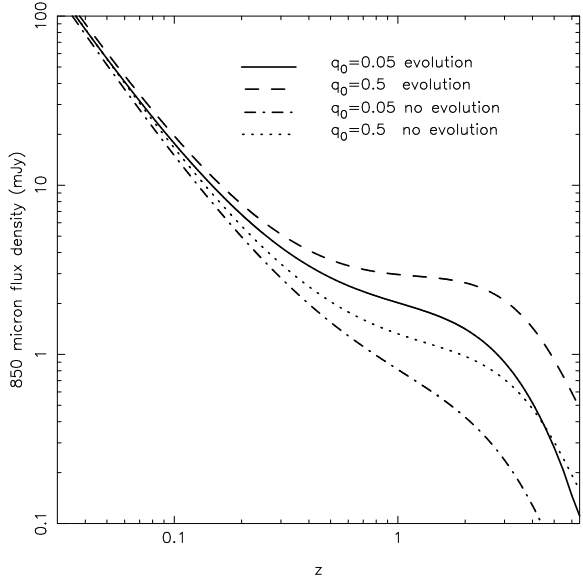}
\caption{If a galaxy has an absolute magnitude $M_B=-22.5$ at the
present day then these graphs show how the received flux from such a
galaxy would vary as a function of redshift using our model with the
parameters described in the previous figure (Fig. \ref{fig:counts}).
The solid line is for a $q_0=0.05$ Universe with Bruzual \& Charlot
evolution, the dashed line for $q_0=0.5$ with evolution, the
dot-dashed line for $q_0=0.05$ without evolution and the dotted line
for $q_0=0.5$ without evolution.  } 
\label{fig:dimming}
\end{figure}

Fig. \ref{fig:counts60} shows our model predictions for the $60\mu m$
differential number counts of IRAS galaxies (Saunders et al. 1990).  This
was an all sky local survey carried out with the IRAS satellite down
to a flux limit of 0.6Jy.  It therefore provides an important test of
our model since spiral galaxies contribute significatly to 
IRAS counts (Neugebauer et al. 1984) 
and so if we are going to assume PLE out to redshifts
of 4 then our local galaxy count predictions at $60\mu m$ need to be 
reasonably consistent with the data.  
The figure shows our evolution and no evolution model(the
$q_0$ makes no difference) and because the IRAS survey was probing
redshifts out to z=0.2 we can see that there is very little difference
between the two models and that they both fit the data reasonably well.
The IRAS counts below 0.2Jy are 
slightly under-predicted using both dust laws, which
could possibly be due to the fact our model doesn't include any fast-evolving 
AGN/ULIRG population. We use the Calzetti dust law with three
dust components of 15, 25, and 32K and this failure of the fainter IRAS counts
is greater than when the 1/$\lambda$ law is used because of the absence
of the 45K dust component, which dominates the thermal emission at $60\mu$m..

We then go on to show in Fig. \ref{fig:counts} our sub-mm predictions
using the Bruzual \& Charlot evolution model with low and high $q_0$
($q_0=0.05, q=0.5$) and also for the corresponding no-evolution models
where we use the Bruzual \& Charlot SED at $z=0$ for all redshifts.
We have used a two-component dust temperature, as described in the
previous section and a galaxy formation redshift, $z_f=4$.  The low
$q_0$ model reproduces the faint counts well, but fails the very
bright counts.  This makes sense since these very luminous sources
would require ULIRG's, having  SFR's of order $\approx$
100-1000M$_\odot$yr$^{-1}$, and/or AGN,  in order to produce
these huge FIR luminosities. Indeed, the $850\mu m$
integral log N:log S appears  flat between 2-10mJy before rising
again at fainter fluxes, suggesting that 2 populations may be contributing to 
the counts.

The high $q_0$ model contains a dwarf elliptical population in order to fit the
optical counts, as already explained, but no dust was invoked in  these
galaxies in the optical models and so they do not contribute to our $850\mu m$
predictions.  Contrary to the optical number counts, the high $q_0$
models predict more galaxies greater than a given flux limit than low
$q_0$ models.  The reason for this is illustrated in Fig.
\ref{fig:dimming}, which shows how the received flux density from a
M$_B=-22.5$ galaxy would vary with redshift in the high and low $q_0$
case, with and without $\tau$=9Gyr. Bruzual \& Charlot evolution.  In the
no-evolution cases the two factors involved are the cosmological
dimming and the effect of the negative k-correction, since we are
effectively looking up the black body curve as we look out to higher
redshift.  The high $q_0$ model( dotted line) predicts greater flux
densities for a given redshift than with low $q_0$, explaining why the
integral number counts are higher for a given flux density.  When the
Bruzual \& Charlot evolution is invoked (solid and dashed lines), we
predict more flux than in the corresponding no-evolution cases at high
redshift, because a galaxy is significantly brighter than at the
present day.  The high $q_0$ model(with evolution) is virtually flat
in the redshift range $0.5 < z < 2$ and the low $q_0$ model again
predicts slightly lower flux densities for a given redshift compared
to high $q_0$. It may be noted that the no-evolution models in this
plot differ slightly from
that of Hughes et al.(1998).  This discrepancy is a result
of the different assumed dust temperature and beta parameter.  
The colder temperature means
that the peak of the thermal emission from the dust is probed at lower
redshifts and so we lose the benefit of the negative k-correction at 
z$\approx$2-3 instead of at z$\approx$7-9 as in Hughes \& Dunlop(1998).

\begin{figure}
\centering
\includegraphics[width=3in,totalheight=3in,angle=-90]{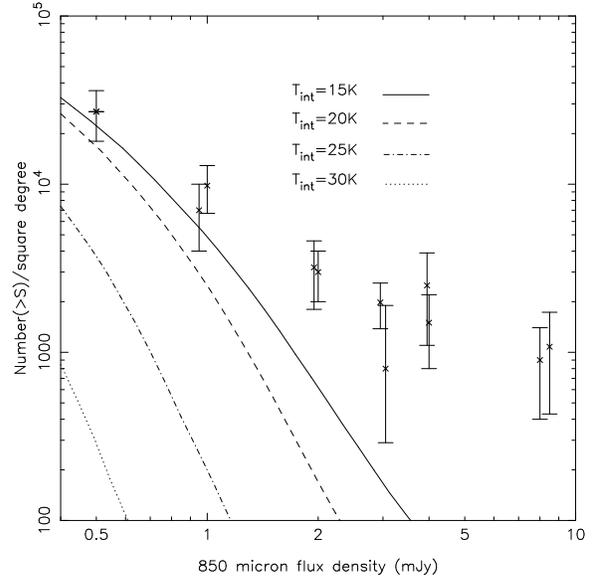}

\caption{The effect of varying the interstellar dust temperature,
$T_{int}$ The graph shows the low $q_0$ model with
$T_{circ}=45K$($\beta=2.0$) and $z_f=4$.  The modest warmer dust
component is included in each plot and the interstellar dust
temperature, which is dominant for the $850\mu m$ counts, is varied
from $15K$(solid curve) to $30K$(dotted curve), again with
$\beta=2.0$.  Our model predictions are sensitive to this variation
and increasing the interstellar dust temperature in fact means we see
less galaxies above a given flux limit.  Typical interstellar dust
temperatures are $\approx 15K$. This trend is perhaps the opposite of what
 you would expect when varying dust temperatures and the reasons are explained
 in section 5. } 
\label{fig:temp}
\end{figure}

Fig.  \ref{fig:temp} shows the effect of altering the interstellar
dust temperature (where we have used the low $q_0$ evolving model).  The
interstellar dust temperature, $T_{int}$ makes a big difference to our
$850\mu m$ number count predictions and the variation is perhaps
contrary to what one may expect in that the lower $T_{int}$ means that
we expect to see more galaxies above a given flux limit S$_{lim}$.
This is because, as we lower the dust temperature, although the
integrated energy ie the area under the Planck curve goes down, the
flux density at $850 \mu m$ goes up slightly because we are seeing the
majority of radiation at much longer wavelengths.  Now recall from the
previous section that the normalisation of the Planck emission curve
is already defined from the amount of flux absorbed by the dust and
the Planck curve is simply scaled accordingly.  So because the
normalisation is fixed, when we lower the dust temperature, we have to
scale the Planck curve up by a much larger factor and therefore find
that we obtain much larger flux densities at $850 \mu m$, explaining
why our models are very sensitive to $T_{int}$.


%
\begin{figure}
\centering
\includegraphics[width=3in,totalheight=3in,angle=-90]{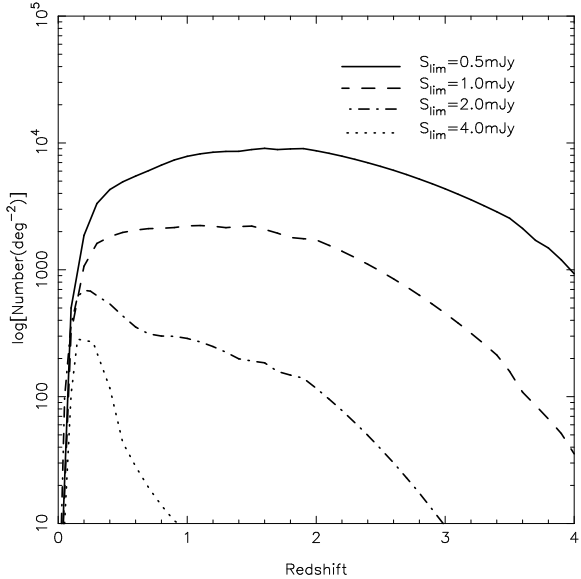}

\caption{The predicted number-redshift distribution of sub-mm selected faint
blue galaxies down to flux limits, $S_{lim}$ of 4.0, 2.0, 1.0 and 0.5mJy.
The graph shows the low-q$_0$ model using the $1/\lambda$ dust law with
the parameters described in Fig. \ref{fig:counts60}  
As the flux limit is increased, the peak in the n(z)
distribution shifts
from around z=1.8 at $S_{lim}$=0.5mJy to much lower
redshifts, reaching z$\approx 0.2$ for $S_{lim}$=4.0mJy.
  } 
\label{fig:n_z}
\end{figure}

We have used a galaxy formation redshift, $z_f=4$ which is reasonable
since sub-mm sources seem to exist out to at least that, but we do in
fact find that adopting $z_f=4$ or $z_f=6$ or indeed $z_f=10$ does not
make any difference to the number counts.  Fig. \ref{fig:dimming}
illustrates this, since at $z > 4$ we are observing radiation that was
emitted beyond the peak of the black-body curve, and so cosmological
dimming is no longer compensated for and all the curves begin to fall
away very quickly explaining why increasing $z_f$ beyond about z=4 makes
essentially no difference to the $850\mu m$ number counts. Of course,
a higher assumed $T_{int}$ would extend this redshift range to beyond z=4.

\begin{figure}
\centering
\includegraphics[width=3in,totalheight=3in,angle=-90]{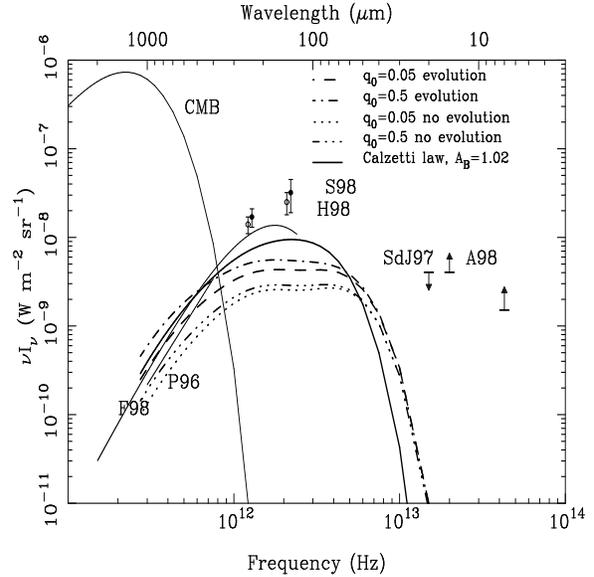} 

\caption{The predicted contribution to the FIR background from our models.  The
latest measurements of the extragalactic FIRB, compared with the COBE
measurement of the cosmic microwave background (Mather et al. 1994). F98 -
Fixsen et al.(1998)(upper solid line) and P96 - Puget et
al.(1996)(lower solid line); H98 - Hauser et al.(1998);
S98 - Schlegel et al.(1998).  Both the Hauser and Schlegel
data each have points at $240\mu m$ and $130\mu m$.  Low and high
$q_0$ models are shown with and without evolution, where we have used
our standard parameters of $T_{int}=15K$, $T_{circ}=45K$, $\beta=2.0$
and $z_f=4.0$ The evolution model, in the low $q_0$ case can account
for all of the FIR background at $850\mu m$, whereas the high $q_0$
one in fact overpredicts by about a factor of 2.  The no evolution
models both underpredict the sub-mm background but are consistent with
it to within an order of magnitude.  The solid curve shows a model
where we have used the Calzetti dust law using $A_B=1.02$ (equivalent to
E(B-V)=0.18 and close to the value 0.15 used by Steidel 
et al for their Lyman Break Galaxies)
 for the dust obscuration 
with a
three-component dust temperature of 15K, 25K and 32K.  It fits the background 
and
faint number counts at $850\mu m$, the IRAS $60\mu m$ counts and also does 
much better in the wavelength range
$100\mu m < \lambda < 500\mu m$. }
\label{fig:background}
\end{figure}

Fig. \ref{fig:background} shows what sort of contribution we get to the
extra-galactic background, simply by integrating over the number
counts in each wavelength bin.  The plot shows the low and high $q_0$
models with and without evolution, and with our standard parameters of
$T_{int}=15K$, $T_{circ}=45K$, $\beta=2.0$ and $z_f=4$.  All the
models predict the same intensity at short wavelengths($\lambda=60\mu$m), 
as low
redshift objects would dominate making the evolution and $q_0$
dependence less significant.  The low $q_0$ model is able to
account for all of the background at $850\mu m$, the high $q_0$ model
in fact overpredicts it by about a factor of 2 and the no evolution
models, although underpredicting it, are still well within an order of
magnitude.  Although we can fit the background at $850\mu m$, we
noticeably fail the data between about $100$ and $300\mu m$.  We find
that the only way to fit these observations using our model is to use
higher values of $A_B$ and higher dust temperatures, as this means
dust is absorbing more energy from each galaxy and so the contribution
to the background in the wavelength range where warmer dust emission
dominates($100\mu m < \lambda < 500\mu m$) is much greater.  The solid
curve in Fig. \ref{fig:background} shows a prediction where we have tried 
the Calzetti dust model which gives more
overall absorption with similar amounts of reddening; this model might
also be expected to fit the B optical counts. We see   that its larger 
amount of absorbed flux allows more flexibility in terms of using more
dust components.  By using three dust temperature 
components 
results we obtain a better (though still not perfect) 
fit to Fig. \ref{fig:background} in the
$100\mu m < \lambda < 300\mu m$ range, while still giving
fits to the IRAS $60\mu m$ (Fig. \ref{fig:counts60}) and faint $850\mu m$
 number counts (Fig. \ref{fig:counts}).

\section{Discussion}

We have taken a different approach from the standard way in which
sub-mm flux's are estimated using UV luminosities \cite{meurer99}.
Instead of assuming a relationship between the UV slope $\beta$ and
the ratio $L_{FIR}/L_{UV}$, we proceed directly from the spiral
galaxy UV luminosity functions and simply re-radiate into the FIR by
assuming a simple dust law constrained from the optical counts.  A
direct result of this, as has already been illustrated in the
previous section, is that decreasing the interstellar dust temperature
actually increases the received flux density at $850\mu m$, firstly
because the peak in the Planck emission curve moves towards longer
wavelengths and secondly because (as the absorbed flux from the dust
is fixed) the normalisation scaling factor goes up.  The fact then
that we model the dust using a dominant interstellar component of
15K, which is significantly colder than that used in models of
starburst galaxies (typically 30-50K), means that we are able to show
that the evolution of normal spiral galaxies like our own Milky Way,
using the Bruzual model
with an exponential SFR of $\tau=9$Gyr, could make a very significant
contribution to the sub-mm number counts in the $S_{850} <
2$mJy range.  Indeed this sort of temperature for spirals has been
given recent support from observations of ISO at $200\mu m$ 
\cite{alton98a} where, for a sample of 7 spirals, a mean temperature
of 20K was found, about 10K lower than previous estimates from IRAS
at shorter wavelengths.  They found that 90 percent of the FIR
emission came from very cold dust at temperatures of 15K.  Sub-mm observations
of spirals (Alton et al. 1998b; Bianchi et al. 1998) and observations of dust
 in our own 
galaxy (Sodroski et al. 1994; Reach et al. 1995; Boulanger et al. 1996; 
Sodroski et al. 1997) also support the claims of these sorts of dust 
temperatures.
Of course, at z=4
our assumed interstellar dust temperature of 15K is comparable to that
of the microwave background. 

Our models show that normal spiral galaxies (ie those that 
evolve into galaxies like our own Milky Way assuming the Bruzual model) 
fail to provide the necessary FIR
flux of the most luminous sources($>2$mJy) and this is not surprising since the
$\tau=9$Gyr SFR at high redshift($z>1$), which is consistent with the UV data,
is lower than that inferred by other  models which fit the sub-mm counts  by a
factor of about 5 or so \cite{blain98a}.  The LBG galaxies at high redshift are
predicted to be evolved spirals by the Bruzual models and the dust we invoke
($A_B$=0.3 implies an attenuation factor at 1500\AA of 2.3) is  enough to make
them low luminosity sub-mm sources at flux levels of around 0.5mJy.  This
amount of dust, though, is not enough to account for the factor of 5
discrepancy and there are several possible reasons for this.  

The first is the possible additional contribution to the sub-mm counts from
AGN.  Modelling of the obscured QSO population has shown that they could
contribute, at most, about 30\% 
of the background at $850\mu m$ but they
can  get much closer to the bright end of the sub-mm number counts
\cite{gunn99}.  This is shown  in Fig. \ref{fig:counts} where we also show the
$q_0$=0.5 model of Gunn \& Shanks. Although the slope of the QSO count at the
faintest limits is too flat, at brighter fluxes the QSO model fits better than
the faint blue galaxy model and the combination of the two gives a better fit
overall.


It is also possible that the optical and sub-mm observations are sampling a
completely different population of galaxies 
as the obscured galaxies sampled
by the sub-mm observations may well just be too red or too faint to be detected
in the UV at the current flux limits 
(Smail et al. 1999, 2000; Dey et al. 1999). That may mean that the most 
luminous
sub-mm sources or ULIRG's($>10^{13}L_{\odot}$) are not the LBG galaxies (which
the Bruzual model predicts as evolved spirals) and so then it would not be
surprising if the current sub-mm and UV derived star-formation histories at
high redshift were different. However, the evidence is growing that the faint
blue galaxies {\it are} significant contributors to the faint sub-mm counts.
Chapman et al.(1999) carried out sub-mm observations of 16 LBG's and found,
with one exception, null detections down to their flux limit of $0.5$mJy. But
their one detection may suggest that with enough SCUBA integration time it
might be possible to detect LBG's that are particularly luminous in the FIR and
indeed, while this paper was in preparation,  work from Peacock et al (1999) 
suggests that faint blue galaxies may be detected at $850\mu m$ at around the
0.2mJy level.  This is below
the SCUBA confusion limit of $\approx2$mJy (Hughes et al. 1998; Blain
et al. 1998b) and highlights
the problem faced by Chapman et al.(1999) in performing targetted
sub-mm observations of LBG's.   The conclusions of Peacock et al.(1999) 
suggest that the LBG population (the faint blue galaxies in our model)
 contribute  
at least 25 percent of the
background at $850\mu m$ and Adelberger et al.(2000) also come to similar 
conclusions, namely that the UV-selected galaxy population could account
for all the $850\mu$m background and the shape of
the number counts at $850\mu$m.  However, the conclusions of Adelberger 
et al.(2000) are based on the fact
that the SED of SMM J14011+0252 is representative of both the LBG and sub-mm
population.   At present, they are only assumptions, but nevertheless 
 the conclusions of all these authors
seem to suggest that ULIRG's may not 
contribute to the faint sub-mm number counts and background as much as 
was first thought.

The spectral slope of the UV continuum and the strength of the H$\beta$
emission line in Lyman Break Galaxies support the fact that interstellar
dust is present 
(Chapman et al. 1999),
but the physics of galactic dust and the way it obscures the optical radiation
from a source is still very poorly understood.  We started by adopting a very
simplistic model for the dust, treating it as a spherical screen around our
model spiral galaxy.  The dust might, in reality, be concentrated in the plane
of the disk for spiral galaxies and may also tend to clump around massive
stars. This would  make the extinction law effectively grayer as suggested by 
observations of local starburst galaxies  (Calzetti, 1997).  Indeed, we have
investigated the effect of the grayer Calzetti extinction law and found that it
would produce a larger sub-mm count contribution due to the higher overall
absorption it would imply. Metcalfe et al (2000) have also suggested that there
may  be evidence for evolution of the extinction law from the  U-B:B-R diagram
of faint blue galaxies in the Herschel Deep Field. 

We have assumed pure luminosity evolution (PLE) throughout this paper. The
assumption that the number density of spiral galaxies remains constant might 
certainly not be the case if  dynamical galaxy   merging is important for
galaxy formation.  However, as we have seen it is relatively easy to fit  the
sub-mm number counts with PLE models whereas it is in fact impossible to fit
the counts using pure density evolution models without hugely overpredicting
the background by 50 or 100 times \cite{blain98a}.  So, if existing sub-mm
observations are correct then although density evolution may also occur,
luminosity evolution may be dominant.  It is also striking how well
the PLE models do in the optical number counts and colour-magnitude diagrams
and together with the fact that we observe highly luminous objects in the
sub-mm out to at least $z=3$ , this could indicate that the biggest galaxies
could have formed relatively quickly, on timescales of about $1$Gyr or so. If
this were true, then the PLE models may be a fair approximation to the galaxy
number density and evolution in the Universe out to $z\approx 3$ in both the
optical/near-IR and FIR.


We have not taken into account early-type galaxies as no dust was invoked in
these in the optical galaxy count models. In particular, we have not included
any contribution from dust in the dE population which is invoked to fit the
faint optical counts in the $q_0=0.5$ model (Metcalfe et al. 1996).
  If we were to include their
possible contribution this would increase our $850\mu m$ counts predictions at
the faint end since in our models both early-type and dE  star formation occurs 
at high redshift which is the region of greatest sensitivity for the sub-mm
counts. At brighter fluxes though, where, in our models, low redshift galaxies
are the only possible influence, the inclusion of early-type galaxies would be
negligible.

\section{Conclusions}

The aim of this paper was to investigate whether, by re-radiating the absorbed
spiral galaxy UV flux into the FIR, the dust invoked in the faint blue
spirals at high z
from the optical galaxy count models of Metcalfe et al.(1996) could have a
significant contribution to the sub-mm galaxy counts and also the FIR
background at $850\mu$m.  We have found that, using a interstellar dust
temperature of $15K$, a modest circumstellar component of $45K$, a beta
parameter of 2.0 and a galaxy formation redshift of $z_f\approx4$ we can account for
a very significant fraction of the faint $850\mu m$ source counts, both in the
low and high $q_0$ cases when we invoke Bruzual \& Charlot evolution (see Fig.
\ref{fig:counts}). These evolutionary models give 5-10 times more contribution
to the faint sub-mm counts than the corresponding no-evolution models. At
brighter fluxes, we find that the SFR and dust assumed in our  normal spiral
model are too low  to produce the FIR fluxes of the most luminous sources.   In
the no-evolution cases, we underpredict the number counts, even at the faint
end.  Our predicted redshift distribution of sub-mm selected faint blue 
galaxies
suggests that the main contribution to the faint counts is in the 
range $0.5 < z < 3$, peaking at $z\approx1.8$.  We have shown that our model 
fits the $60\mu m$ IRAS data well, an
important local test if we want to assume PLE and extrapolate our optical
spiral galaxy luminosity functions out to higher redshift. With the evolution
models we can easily account for 50-100\% of the FIR background at $850\mu m$ 
but fail the data by nearly an order of magnitude in the $100-300\mu m$ range. 
We have shown that the only way to fit these observations using this optically
based model is to use assume more dust obscuration ($A_B=0.6$) and much warmer
dust (T=30K).  Effectively gray extinction laws such as that of Calzetti et al
(1997) may also provide more overall absorption and hence allow more dust
temperature components to allow the flexibility to fit  the FIR background from
60-850 $\mu m$. However, the bright  sub-mm counts will  still require a
further contribution from QSO's or ULIRGs to complement the contribution of the
faint blue galaxies at fainter fluxes.

\end{document}